\documentclass[12pt,a4paper]{article}
\usepackage{amsmath}
\usepackage[dvips]{graphicx}
\begin{document}

\title{ 
\vspace{2cm}
Hierarchical clustering and formation of power-law correlation
in   1-dimensional self-gravitating system
}

\author{
  Hiroko KOYAMA\thanks{\tt hiroko@allegro.phys.nagoya-u.ac.jp} \, 
and Tetsuro KONISHI\thanks{\tt tkonishi@allegro.phys.nagoya-u.ac.jp} \\
Department of Physics, School of Science,\\
Nagoya University, 464-8602, Nagoya, Japan
}
\date{August 31, 2000}

\maketitle

\begin{abstract}
The process of formation of fractal structure  in 
one-dimensional self-gravitating system is 
examined numerically.
It is clarified that 
structures created in small spatial scale grow up to 
larger scale through clustering of clusters, and form power-law
correlation. 
\end{abstract}

\section{Introduction}

1-dimensional self-gravitating system has long been used to 
study systems interacting through 
self-gravity~\cite{oort-sheet,hohl-feix-1967,Miller-1,yamashiro-gouda-sakagami,sheet-tgk-2}. The model helped us to understand various properties,
such as motion of starts normal to the galactic plane,
relaxation properties of elliptical galaxies, and so on.

Recently we have found that fractal structure dynamically emerges
in the system~\cite{hktk-1}.  
It is surprising that the fractal structure is  created
from non-fractal initial conditions. 

The emergence of fractal structure has several important meanings.
First, there are many fractal (or power-law type) structure 
in nature. In  astrophysics the distribution of 
young stars,  galaxies and  cluster of galaxies are known to obey 
power-law~\cite{Larson,Hanawa-cluster,hypergalaxy-power}.
However there is no universal explanation about the origin of 
power-law behavior in general.
Since the dynamical emergence  of fractal structure
arises in  self-gravitational model,
there can be a relation between the origin of these fractal structures.

Second.
There are many studies about the relaxation process
of 1-dimensional self-gravitating system
and it is reported that relaxation is quite
slow (\cite{Miller-1,sheet-tgk-2}, and references therein).
It is interesting if there is any relation between the 
 slow relaxation in the model and  
the dynamical emergence of fractal structure.

Since the  system is Hamiltonian system with many particles,
one may expect that most initial condition lead the system to 
thermal equilibrium. 
It is interesting that
some initial conditions,
e.g., uniformly random in spatial distribution, 
evolves to form fractal spatial structure,
instead of just become thermalized. 
Hence to study the origin of the dynamical emergence of fractal structure
may give unique insight
to non-thermal behavior and structure formation in 
many-body systems in general~\cite{tk-cluster,onion}.

Therefore it is quite interesting to investigate the reason
why the fractal structure emerges from non-fractal initial conditions
in 1-dimensional self-gravitating system. 
In order to do that, it is essentially important to know in detail 
the process how the fractal
structure is created, not just to observe the grown-up fractal structure.

%

In this letter we continue the study 
to investigate the process how the fractal structure
is developed through time evolution.

The model is presented in section 2. In section 3 we will  examine 
in detail how the power-law correlation is formed, and the final 
section is for summary and discussions.

\section{Model}
The model we use in this letter 
is the same as the one used in our previous letter~\cite{hktk-1}, that is,
the one-dimensional self-gravitating system.
The Hamiltonian is
\begin{eqnarray}
  H &=& \sum_{i=1}^N \frac{p_i^2}{2m} + 2\pi G m^2\sum_{i > j} \left|
x_i - x_j\right| \  , \ \ -\infty < x_i < \infty \ . 
  \label{eq:sheet-hamiltonian} 
\end{eqnarray}

This model represents one-dimensional motion of parallel sheets of infinite 
extent interacting through  Newtonian gravity. 

The model is simple and tractable, and yet it also
contains essential feature of gravitational interaction.

Throughout this paper we set 
\begin{equation}
  m \equiv 1/N \ \ \text{and}  \ \  4\pi G \equiv 1 \  .
\end{equation}

\section{Formation of power-law correlation}
In the previous letter~\cite{hktk-1} we showed that fractal structure
emerges from non-fractal initial conditions. Typical initial conditons
we used are those of virial ratio $2E_{kin}/ E_{pot} = 0$.
(Spatial distribution is set to be random.) These state of zero
velocity dispersion corresponds to the limiting case of zero 
thermal fluctuation.

To analyze the spatial structure we use two-point correlation function
$\xi(r)$, which is defined as
\begin{equation}
dP = n dV (1 + \xi(r))
  \label{eq:2-body-correlation}
\end{equation}
where $dP$ stands for probability to find another particle 
in volume $dV$ at distance $r$ from a particle. $n$ is the 
average number density. If particles are distributed independently 
and no spatial structure is formed, then $dP$ does not depend on
$r$ and $\xi(r)=0$. If, on the other hand, $\xi(r)$ increases as
$r$ decreases, the probability that we find another particle
near one particle gets large, and we can understand cluster-like
structures are formed. If the $\xi(r)$ does not have characteristic 
length, then we can understand that the cluster system
does not have characteristic length scale.

\begin{figure}[hbtp]
  \begin{center}
    \includegraphics[width=7cm]{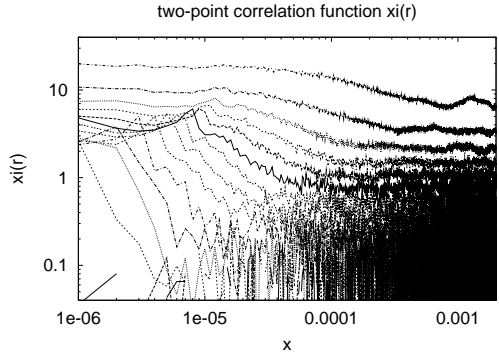}
    \caption{Evolution of correlation function $\xi(r)$
      at $t=\frac{5}{64} \ell$ , $\ell = 3,4, \cdots, 17$
      (from bottom to top).
      System size $N=2^{14}$, Initial condition: $x_i =$random,
      $v_i=0$. $E_{total} = 0.25$.
}
    \label{fig:corr-multi}
  \end{center}
\end{figure}

Fig.\ref{fig:corr-multi} shows temporal evolution of 
$\xi(r)$  for early stage of time.
We can clearly see how the power-law behavior
of correlation function grows up. That is, power-law structure
is first created in small scale, then gradually grows up
to larger scale.

The graphs are aligned from bottom to top as time proceeds
because we are looking at small-sized scale here ($r<0.002$ )
and particles initially separated are coming near,
which increase $\xi(r)$ for small $r$. 

The time dependence of two point correlation function $\xi(r)$ 
tells us when and where the power-law structure is created and
how it grows up. However  $\xi(r)$  does not tell what kind of 
structure is actually formed. So we need another quantity to 
understand the detail of the structure formation.

Since we are dealing with a gravitational system, interaction is 
attractive. 
Each particle, initially 
distributed in the 1-dimensional space, 
attracts each other by gravitational force,
and small clusters are formed here and there.
These clusters attracts and  moves around each other,
as well as their internal oscillation (rotation, in 3D). 
While these processes go on,
the order of 
particles given at initial condition is exchanged, and 
initial distribution, aligned on the $x$-axis as seen in the 
$(x,v)$ phase space ($\mu$-space), get folded many times.
Hence the degree of folding is a good quantity to understand 
how the cluster structure is formed.

We can measure the number of folding as follows.
Since we are considering the case where initial conditions have
zero velocity dispersion, distribution of particles 
in $(x,v)$ space ($\mu$-space) in each initial condition is on 
a curve, and we can label particles along the curve.
For example, if we take the initial conditon as 
\[
v_i(t=0) = 0 , \ \  \  i = 1,2, \cdots, N \ , 
\] 
we can label the particles as
\begin{equation}
x_1(0) < x_2(0)  < \cdots < \ x_N(0) \
\label{eq:particle-order}
\ .
\end{equation}
 
Then we can count $f(t)$, the number of folding at time $t$,  as
  \begin{align}
    \label{eq:folding}
f(t)& \equiv 
\text{ number of  \ $i$ \ $ (1 < i < N) $ which satisfies}\nonumber\\
&\left(x_{i+1}(t) - x_{i}(t)\right) \cdot\left(x_{i}(t) - x_{i-1}(t)\right) < 0
  \end{align}

\begin{figure}[hbtp]
  \begin{center}
\includegraphics[width=8cm]{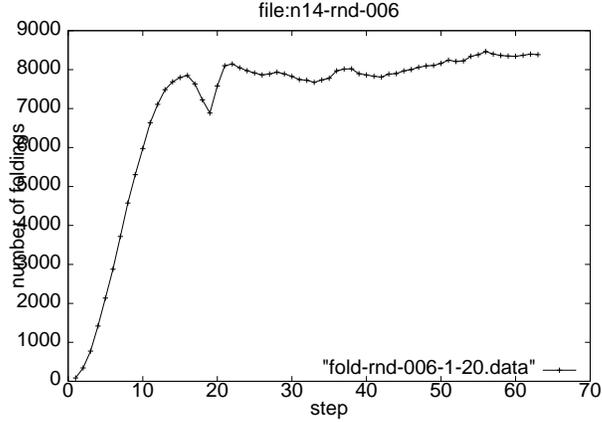}
    \caption{Time evolution of number of foldings.
      Horizontal axis represents time measured in 
      steps $\ell$ defined in 
      eq.(\protect\ref{eq:steps_and_time})}
    \label{fig:folding-times}
  \end{center}
\end{figure}

Fig.~\ref{fig:folding-times} shows how the initial distribution 
becomes folded. 
Time dependence of number of folding $f(t)$ shown in this figure
and 
time dependence of the two point correlation function $\xi(r)$ 
shown in Fig.~\ref{fig:corr-multi} 
 corresponds well. 
That is, power-law behavior of $\xi(r)$ develops from small scale as
the number of folding increases, and become most clear when
$f(t)$ is at a maximum at $t=t_{16}$.


The process of structure formation, as seen in two quantities
$\xi(r)$ and $f(t)$, can also be understood in the real space of 
particle distribution.  Fig.~\ref{fig:fold-08-16},
 represents time evolution of particle distribution
of particles $(x_i(t),v_i(t))$ , $8200 < i < 8300$ for
$t=t_8, t_{13},t_{15}$, and $t_{16}$,
where $t_\ell$ are time steps defined as
\begin{equation}
  t_\ell \equiv \frac{5}{64} \ell \ .
  \label{eq:steps_and_time}
\end{equation}
(These time steps are defined just for  practical purpose.)

In these figures we connect each particles on $(x,v)$ space
according to the order defined at the initial 
condition~(\ref{eq:particle-order}) so as to enhance the 
structure of folding. 
Vertical steps on the graphs represent the location where particles 
are clustered.

At $t=t_8$ we can see several small clusters are formed.
These clusters gradually approach each other to form
larger clusters, as seen in $t=t_{13}$, $t_{15}$ and $t_{16}$
in the figure.  Through this process power-law correlaion
is developed.

\begin{figure}[hbtp]
    \includegraphics[width=7cm]{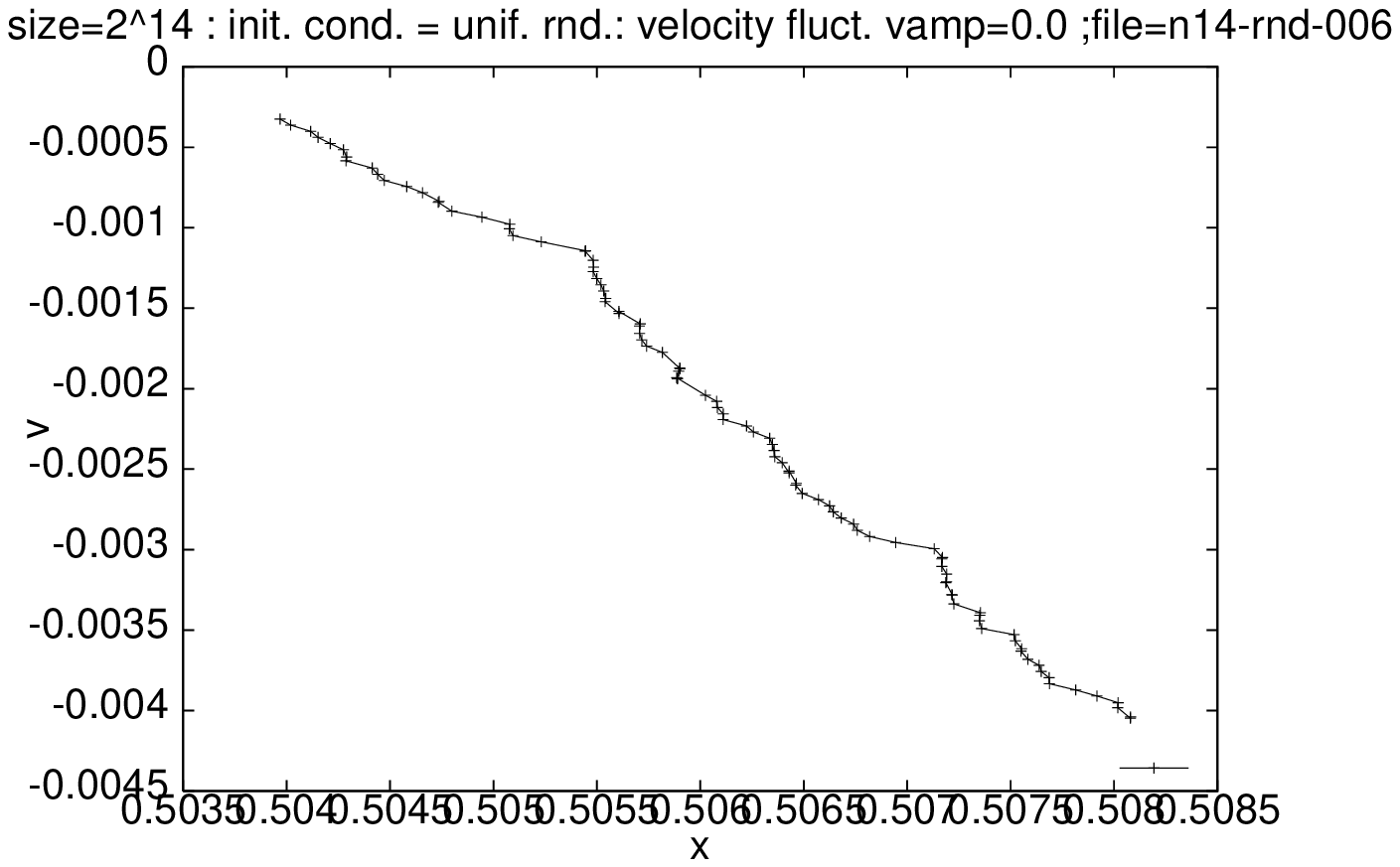}
    \includegraphics[width=7cm]{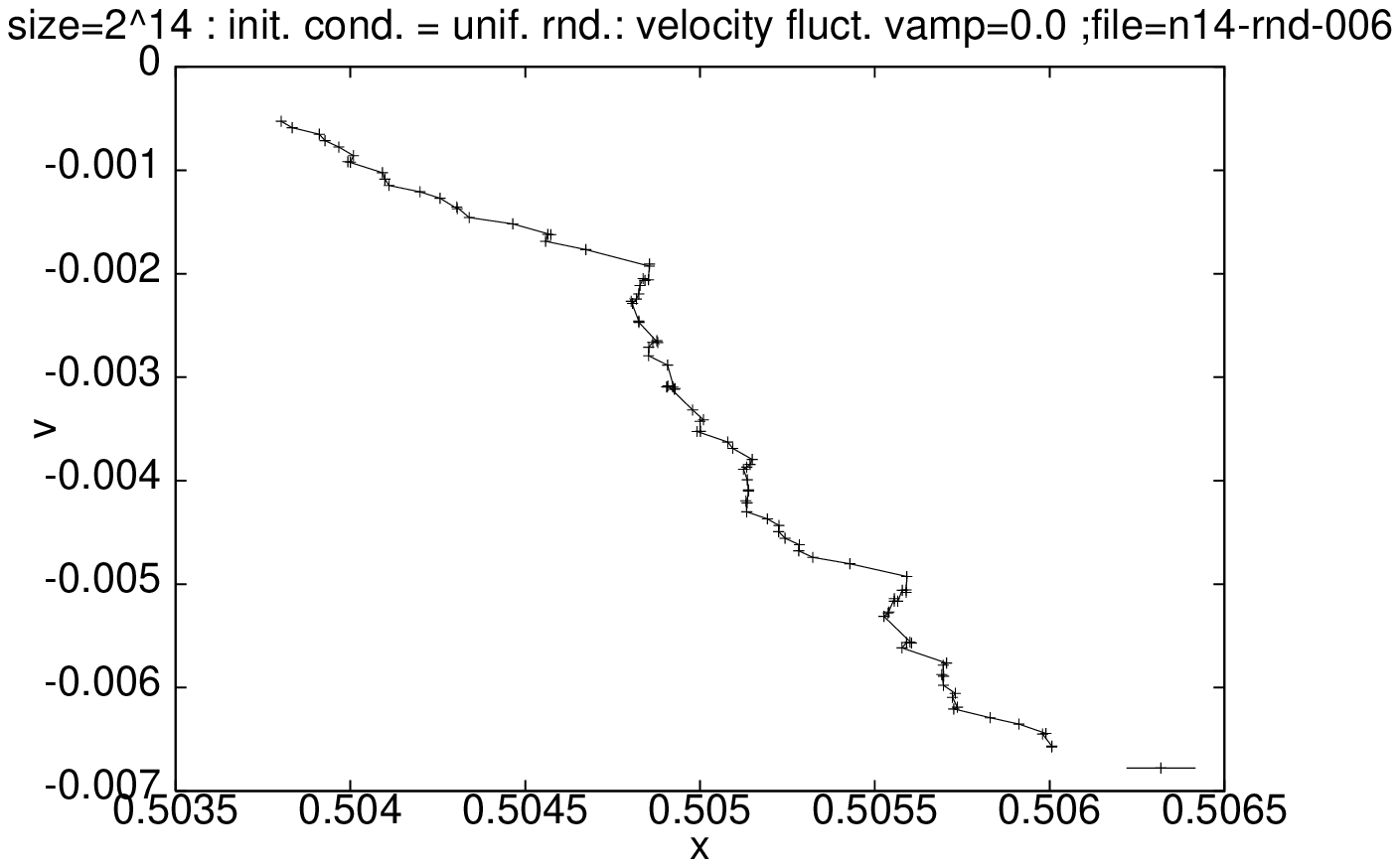}
    \includegraphics[width=7cm]{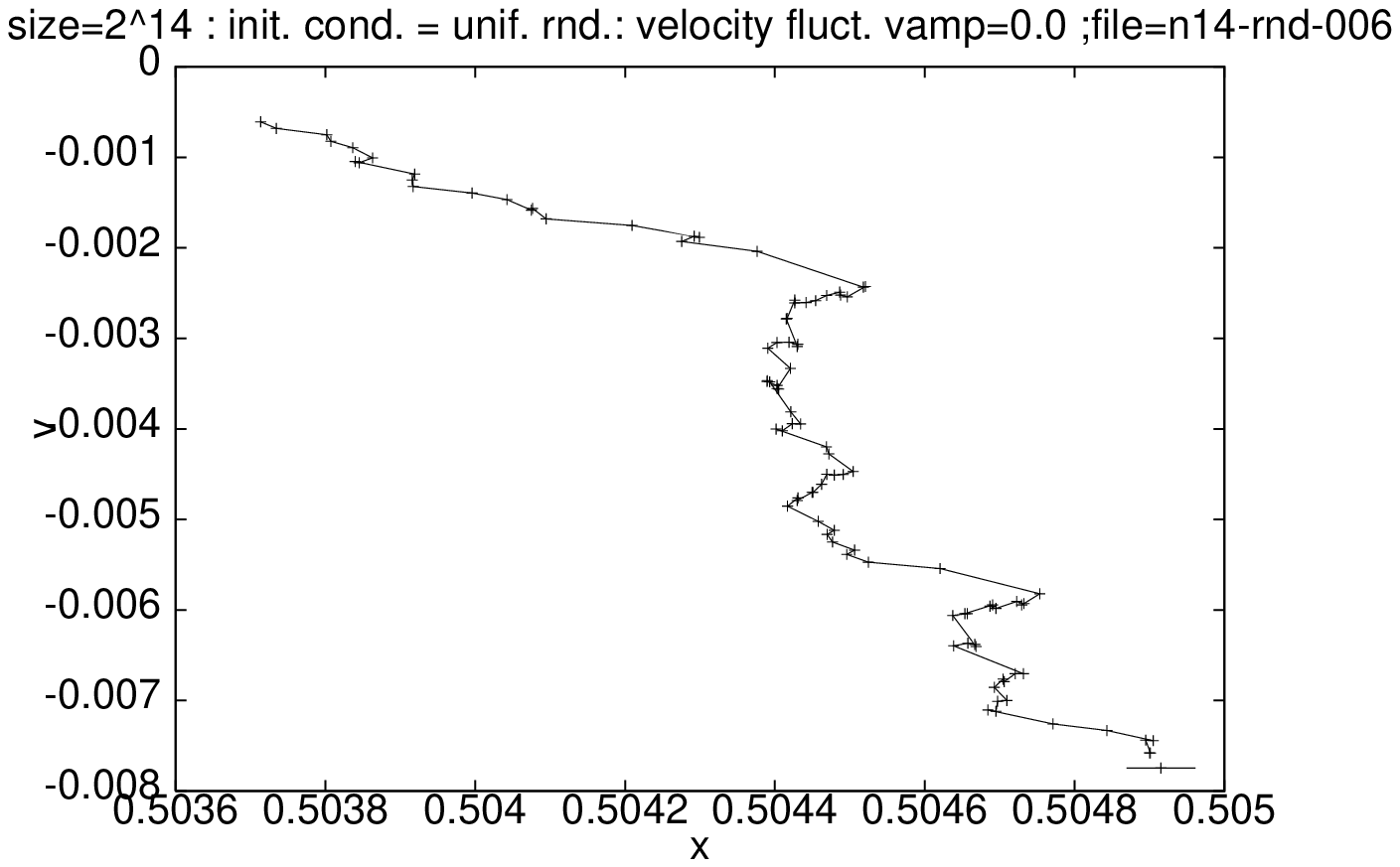}
    \includegraphics[width=7cm]{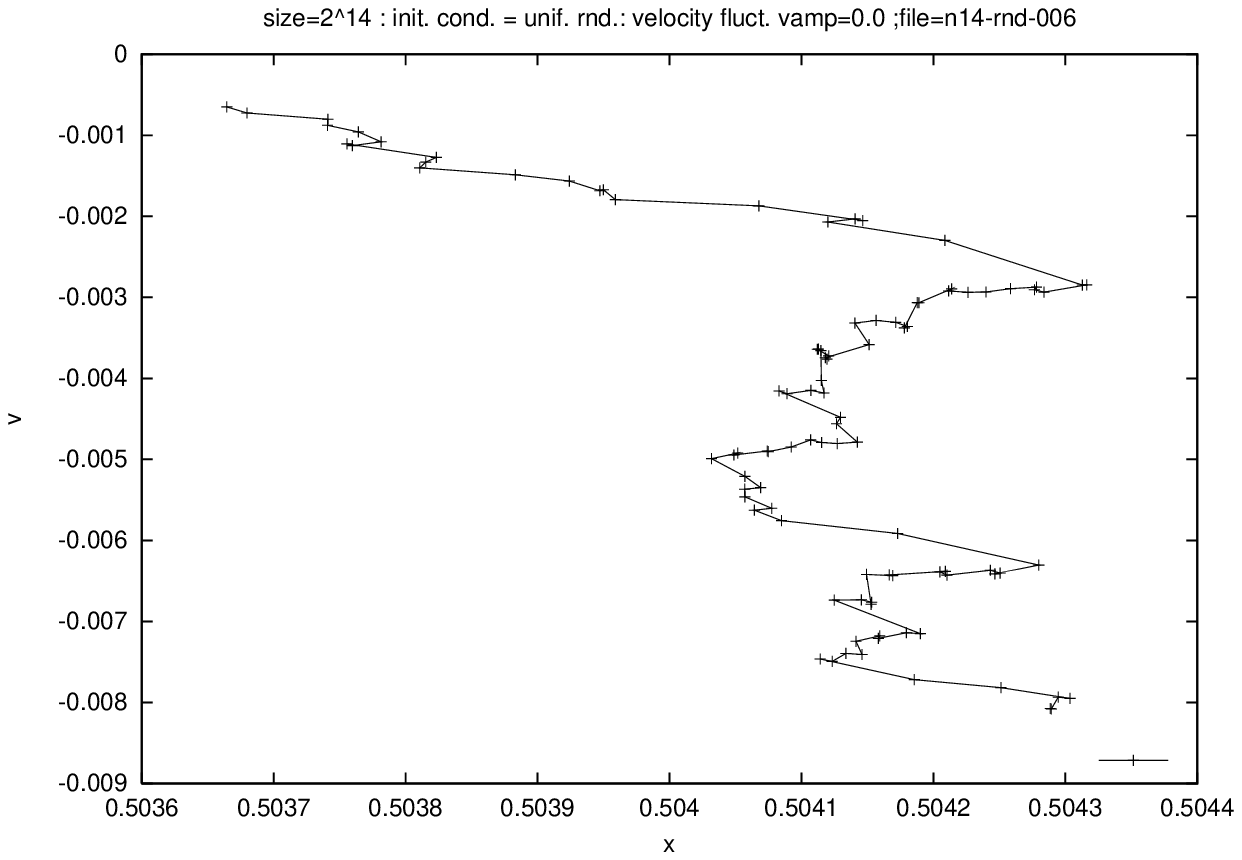}
    \caption{Distribution in 
      $(x,v)$ space at $t=t_\protect{8}$ (up:left),
      $t_{13}$(up:right), $t_{15}$(down:left)
      and $t_{16}$ (down:right) for particles  No.
      $8201\sim 8299$ . Particles are linked according to the order
      assigned at $t=0$.}
    \label{fig:fold-08-16}
\end{figure}

In later time the  clusters get together and larger clusters
are developed. In the left of Fig.~\ref{fig:fold-55} 
we show a snapshot of the distribution of particles 
shown in the previous figure Fig.\ref{fig:fold-08-16}.
This cluster itself is a part of a larger cluster system
shown in the right of the figure.

\begin{figure}[hbtp]
    \includegraphics[width=7cm]{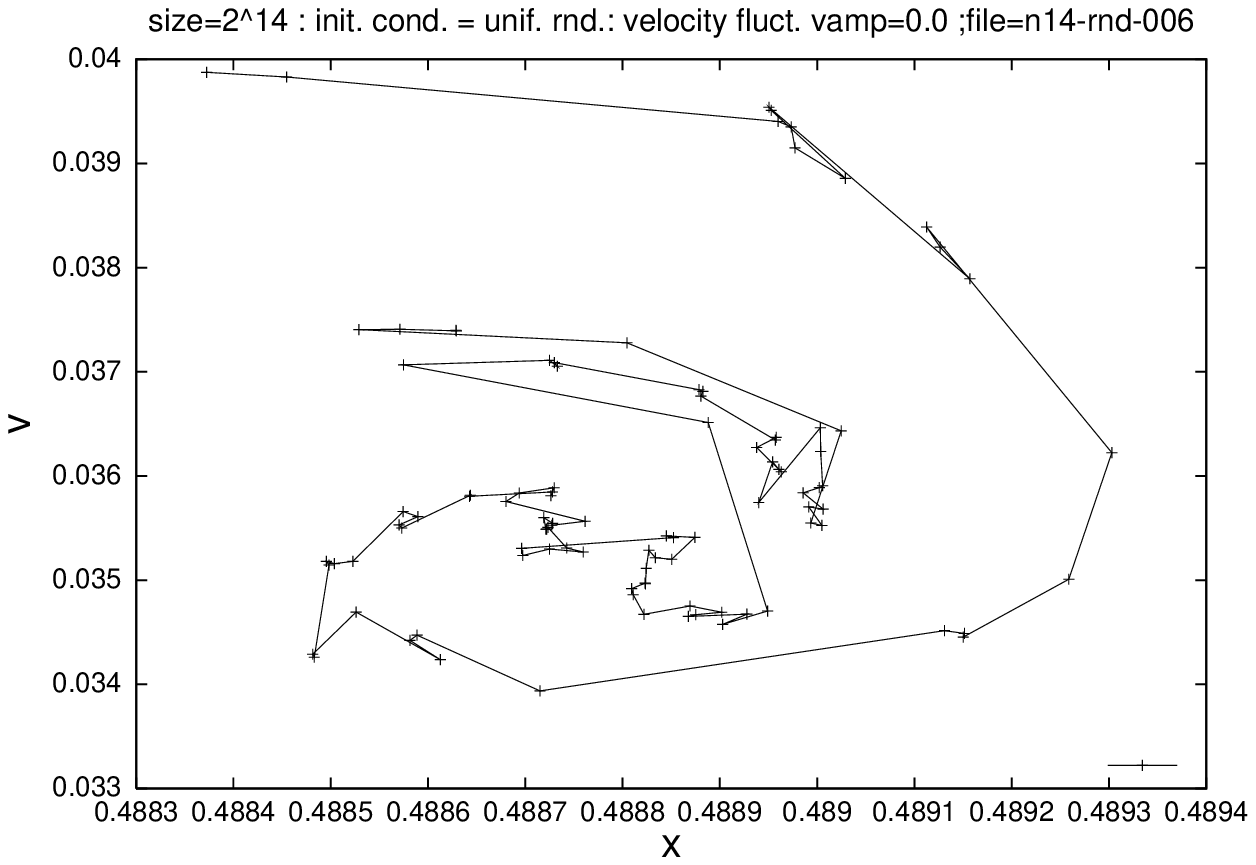}
    \includegraphics[width=7cm]{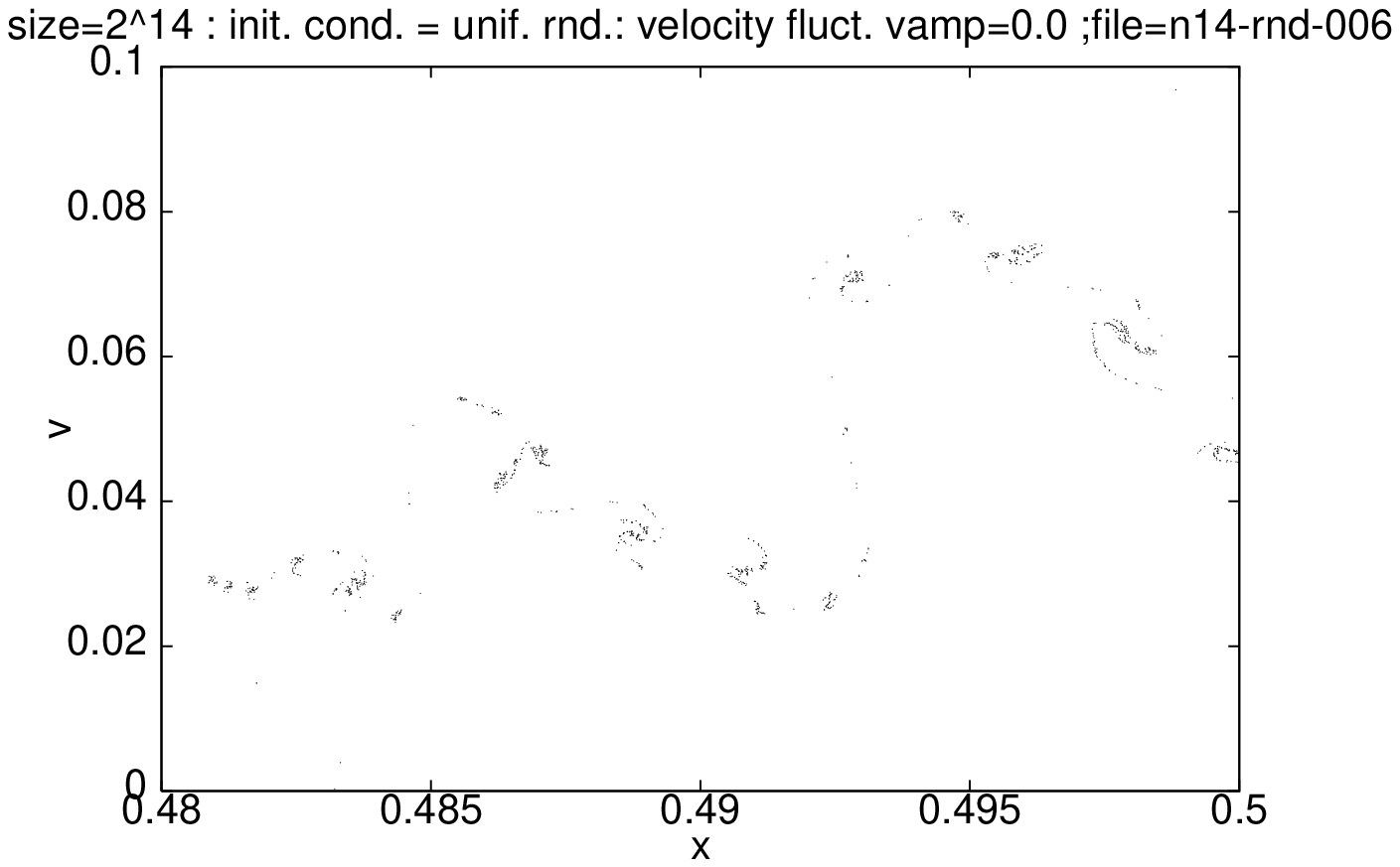}
    \caption{continued from Fig.\protect\ref{fig:fold-08-16} (left)
            and $\mu$-space (wide ranged) (right) at $t=t_\protect{55}$ }
    \label{fig:fold-55}
\end{figure}

\section{Summary and discussions}
In this letter we clarified the process of formation of
power-law correlation in one-dimensional self-gravitating system,
found in our previous letter~\cite{hktk-1}.
Power-law behavior of two-point correlation function 
begins in small spatial scale, then  grows up to larger 
scale. Number of foldings in the order of particles
gets large when the power-law behavior is clearly seen.
These results and phase space portraits represents the process of
formation of fractal structure as follows:
tiny  clusters are formed first, then the clusters 
get together to form larger clusters, and so on. 

The process we showed is quite similar to the ``hierarchical clustering
scenario''~\cite{peebles-book,mandelbrot}, 
which is used for descibe the large-scale structure of universe. 
It is known that two-point correlation function of 
galaxies and cluster of galaxies obey power-law~\cite{hypergalaxy-power}.
An explanation for the origin of the power-law structure is that
density fluctuation of small scale is created first 
to become some structure of small scale (bottom-up), 
and, according to the ``hierarchical clustering scenario''
the structure made in small scale attract each other through 
gravitational interaction
and merge into larger structure, and so on. 
The initial state is now considered to be of 
small temperature fluctuation. 

In this letter we have shown one example which obeys this
``hierarchical clustering scenario'' by purely gravitational interaction.
An important fact is that our result shows that power-law correlation
emerges through hierarchical clustering.
Also the initial condition we found for fractal structure to be created
is the zero velocity dispersion, which   coinsides with zero temperature
fluctuation considered in the  cosmology.

The clustering process is developed by gravitational interaction.
Hence similar process can be  actually occuring in nature where
the dynamics of the system is governed by gravity. One candidate is 
the power-law distribution of yound stars in star-forming
regions~\cite{Larson,Hanawa-cluster}.

The emergence of power-law correlation can be 
considered as a subject of 
structure formation in general, 
Our result shows a process of generating fractal structure by
purely dynamical interaction. 

In this letter we have clarified how the fractal structure is formed
from non-fractal initial conditions in one-dimensional  self-gravitating 
system. We hope that this result helps us to understand why the structure
is created.


\newcommand{\etalchar}[1]{$^{#1}$}

\end{document}